# TITLE

Massive viral replication and cytopathic effects in early COVID-19 pneumonia

## AUTHORS

L. Schifanella[1]*†, J.L. Anderson[2]†, M. Galli[3,4], M. Corbellino[3], A. Lai[4], G. Wieking[2], B. Grzywacz[5], N.R. Klatt[1], A.T. Haase[6]* and T.W. Schacker[2]*

## AFFILIATIONS

[1] Division of Surgical Outcomes and Precision Medicine Research, Department of Surgery, Medical School, University of Minnesota, Minneapolis, MN 55455, USA.

[2] Department of Medicine, Medical School, University of Minnesota, Minneapolis, MN 55455, USA.

[3] III Division of Infectious Diseases, ASST Fatebenefratelli Sacco, Milan, Italy

[4] Department of Biomedical and Clinical Sciences Luigi Sacco, University of Milan, Milan, Italy

[5] Department of Laboratory Medicine and Pathology, Medical School, University of Minnesota, Minneapolis, MN 55455, USA.

[6] Department of Microbiology and Immunology, Medical School, University of Minnesota, Minneapolis, MN 55455, USA.

†These authors contributed equally to this work.
*Corresponding authors

## FULL SUMMARY PARAGRAPH

SARS-CoV-2 is the cause of COVID-19 acute respiratory illness that like its predecessors, MERS and SARS, can be severe and fatal [1-4]. By April of 2020, COVID-19 infections had become a worldwide pandemic with nearly 3 million infections and over 200,000 deaths. The relative contributions of virus replication and cytopathic effects or immunopathological host responses to the severe and fatal outcomes of COVID-19 lung infections have as yet to be determined. Here we show that SARS-CoV-2 replication and cytopathic effects in type II alveolar pneumocytes causes focal lung injury in an individual with no history of pulmonary symptoms. These findings point to the potential benefit of early effective antiviral treatment to prevent progression to severe and fatal COVID-19 pneumonia.

**MAIN TEXT**

Understanding the mechanisms responsible for the COVID-19 fatal lung infections is important to design potentially life-saving treatments. The virus itself, an immunopathological response to infection or a combination of these direct and indirect mechanisms could be responsible for lung injury. The distinction is important, because antiviral drugs and antibodies would be appropriate to treat lung injury largely due to virus replication and cytopathic effects if direct virus insult is the major driver of lung pathology. Clinical trials underway to test antiviral drugs are designed to determine if the inhibition of SARS-CoV-2 replication can prevent or mitigate the pathological consequences of lung infection and thus far these trials are aimed at people with symptomatic and often advanced disease. On the other hand, if severe pneumonia is due to an immunopathological response to the virus, then moderating that response, e.g., by inhibiting IL-6 with the monoclonal antibody tocilizumab or other immunomodulatory approach, would be appropriate, but with the risk, as in any immunosuppressive treatment, of impairing the immune response to infection.

Ground glass opacities or patchy infiltrates in the lungs in CT images of asymptomatic SARS-CoV-2 infected individuals[5-7] suggest that lung infection and associated tissue injury may be detectable in individuals who did not have severe pneumonia or respiratory failure[7]. Studies of tissues at this stage might therefore provide a glimpse of the relationship between viral replication and cytopathic effects to lung injury before that relationship becomes blurred by progression to end stage pneumonia. We had an opportunity to test this hypothesis in analysis of lung specimens obtained postmortem from an 88 year-old Italian woman who succumbed to a heart attack while quarantined in a hotel after testing positive for SARS-CoV-2. She is described as afebrile and in good health without cough, shortness of breath or any other symptoms of clinically manifest SARS-CoV-2 infection. We describe findings in the lung of this individual at a stage of infection without manifest pneumonia or respiratory failure that support direct virus-mediated tissue destruction in the pathogenesis of COVID-19 pneumonia.

**Focal COVID-19 pneumonia**

We received paraffin embedded lung specimens from the SARS-CoV-2 infected subject of this report from the Department of Pathology of Fatebenefratelli Sacco Hospital, Lombardy, Milan, Italy. H&E stained lung tissue showed heterogeneous organ damage **(Extended Data Fig.1)**. Alveolar structure was partially conserved, although emphysematous changes were clearly recognizable in the region defined by the blue box and arrow **(Extended Data Fig.1)**. In the region defined by the red box and arrow, there is a very well delimited area of striking consolidation. At higher magnification there is evidence of diffuse alveolar damage with exudative characteristics that include hyaline membranes, inflammatory cell infiltrates (mainly lymphocytes), desquamated type II pneumocytes, microvascular thrombosis, and vascular endothelial cell damage with red blood cells in the alveolar spaces. We focused our analysis on this region of lung injury. We used RNAscope in situ hybridization (ISH) with anti-sense SN probes to detect SARS-CoV-2-specific Spike (S) region genomic sequences, and nucleocapsid (N) region genomic sequences shared with SARS- and MERS-CoV, to detect viral genomic RNA+ (vRNA+) in infected cells. We first show images consistent with spread of infection into the lung by infection of the bronchial epithelium **(Fig. 1)**. Degenerating vRNA+ cells line a small and larger bronchiole **(Fig. 1A,B, short arrows)** and a grape-like cluster of vRNA+ hyperplastic epithelium we later show to be type II pneumocytes is adjacent to infected bronchiolar epithelium, consistent with this mode of spread into alveoli. There are also vRNA+ clusters adjacent to the alveolar space **(Fig. 1A, long arrows)**, vRNA released from lysed type II pneumocytes **(Fig. 1C, red arrow)** and CD68+ macrophages that appear to have phagocytosed vRNA+ cells and debris **(Fig. 1C brown and red arrows)**. Detection of SARS-CoV-2 RNA was specific because 1) there was no signal when the SN probe was hybridized to normal lung tissue (**Extended Data Fig.2A**); 2) a negative control probe did not detect SARS-CoV-2 RNA (**Extended Data Fig.2B**) in the lung tissues from the same positive focal area shown in **Fig. 1**; 3); and the SARS-CoV-2-Spike specific probe alone (**Extended Data Fig.2C**) detected vRNA at similar levels to the SN probe mix, which would also hybridize to SARS-CoV and MERS RNA (**Extended Data Fig.2D**).

**Massive virus production, spread and cytopathic effects of SARS-CoV-2 infection in type II pneumocytes**

We identified type II pneumocytes as the principal cells in which SARS-CoV-2 replicates and produces virus. To do so, we combined ISH with Napsin A staining to detect vRNA in Napsin A+ type II pneumocytes and ISH with TSA/ELF amplification[8] to render RNA in virions visible for light microscopy to score for virus producing cells. With these two methods, we could then examine the spread of infection and cytopathic effects in the lung. In the numbered regions in the overview of infection and spread **(Fig. 2)**, there are foci or collections of vRNA+/Napsin A+ type II pneumocytes in close spatial proximity to one another and to vRNA negative Napsin A+ type II pneumocytes **(Fig. 2, regions 1-5)**. These images suggest a chain of transmission events from an infected cell to adjacent susceptible cells. We will return to the role of macrophages in phagocytosis of infected cells and virus released from dying cells but note here that a RNA+/Napsin A- non epithelial cell overlies extracellular viral RNA+ derived from lysis of infected cells **(Fig.2 region 6)** and a conjugate of viral RNA+/Napsin A+ and RNA+/Napsin A- non epithelial cell **(Fig.2. region 7)**. Both images are consistent with subsequent documentation of macrophage acquisition of viral RNA by phagocytosis.

The visualization of virus production, spread and cytopathic effects in the lung confirmed the predominant focal nature of spread **(Fig. 3A)** in clusters or chains of susceptible cells in close spatial proximity **(Fig. 3B)**. The image at higher magnification **(Fig. 3C)** of individual virions, virus aggregates and inclusion bodies in and around dying cells captures the massive production of virus and associated cytopathic effects. This extraordinary level of virus production, fusion and cell lysis leaves a visible record in the lung of vast swaths of lysed virus + cells surrounding airways, which are themselves lined by fused virus + cells, and around rare still recognizable virus producing cells **(Fig. 3D)** as well as mats of virions from necrotic type II pneumocytes **(Fig. 3E)**. These images thus support the mode of spread into and within the lung and underscore the extent of tissue injury directly attributable to SARS-CoV-2 replication and cytopathic effects.

**Macrophages acquire vRNA+ by phagocytosis**

With these images of the massive cytopathic effects and cell lysis in mind, we return to the question of whether vRNA+ macrophages represent cells in which SARS-CoV-2 is replicating or macrophages that acquire vRNA by phagocytosis. We have already shown images that suggested that vRNA+ macrophages could acquire vRNA by phagocytosis of infected type II pneumocytes or vRNA released from lysed cells **(Fig. 1, 2)**, and now show further evidence in support of that conclusion. In the region enclosed by a rectangle adjacent to vRNA+ lysed epithelium **(Fig. 4A)** shown at higher magnification **(Fig. 4B)**, a CD68+ vRNA+ contacts lysed

vRNA+ lysed epithelium. The white line outlines the RNA+ positive area of the macrophage that is in contact with the viral RNA in the necrotic cells, suggesting phagocytic acquisition of vRNA. Small CD68+vRNA-negative cells are also in cell-to-cell contact with CD68-negative vRNA+ type II pneumocytes, consistent with a sequence in which macrophages will acquire RNA from dying infected cells. Moreover, macrophages with little vRNA also overly sheets of lysed vRNA+ cells consistent with acquisition of vRNA from infected pneumocytes at an advanced stage of infection in these cells (**Fig. 4C, D**). This conclusion is further supported by the co-localization of vRNA inclusions in LAMP1+ lysosomes in CD68+ cells in some cases within CD68+ extensions of a macrophage (**Fig. 4E, F**).

**IL-6 production by type II pneumocytes**

We investigated what local production of IL-6 might contribute to the inflammatory component of lung injury by staining tissues for IL-6. The morphology and geographical distribution of the IL-6 + cells suggested that the IL-6+ cells were hyperplastic infected type II pneumocytes (**Extended Data Fig.3A**), and we confirmed their identity by showing that the IL-6+ cells were vRNA+ and Napsin A+ (**Extended Data Fig.3B**). While IL-6 production by infected cells could contribute to immunopathology, local production could also be involved in macrophage recruitment and epithelial reparative processes[9-13].

**Conclusion**

We show in a SARS-CoV-2 infected individual with ostensibly early infection, who had no known history of pulmonary symptoms, that there was a region of the lung with focal pneumonia in which massive SARS-CoV-2 replication and cytopathic effects in type II alveolar pneumocytes directly contributes to lung pathology. The extent of diffuse alveolar damage from the lytic effects of infection of type II pneumocytes could explain the current observation that many people testing positive for SARS-CoV-2 have significant hypoxia often in the absence of overt symptoms of pneumonia. It might therefore be prudent in this stealth phase of SARS-CoV-2 lung infection to use pulse oximetry to measure blood oxygen levels as a guide to early supplemental oxygen. Our findings further point to the potential therapeutic benefit of early treatment to inhibit viral replication to prevent progression to severe and fatal pneumonia.

**MAIN REFERENCES**

**METHODS**

**RNAscope in situ hybridization/ Immunohistochemistry[1]**

SARS CoV2 RNA was visualized by RNAscope in situ hybridization with anti-sense probes and reagents from Advanced Cell Diagnostics. Briefly, slides were boiled in RNAscope Pretreat citrate buffer for 15 minutes. Pretreat 3 reagent 3 (protease solution) was added at a 1:15 diution, and the slides were incubated for 20 minutes at 40 F in the hybridization oven. SARS-CoV-2 antisense probes VnCoV2019S 21631-23303 of NC 045512.2, which hybridizes specifically to the 5' end of SARS-CoV-2-Spike RNA, and VnCoV-N 28275-29204 of MN908947.3, which cross-hybridizes to SAR-CoV and MERS N-RNA, were added for 2 hours before continuing with AMPs 1-6 from the RNAscope 2.5 Red detection kit. Warp red chromogen (Biocare Medical) was added to visualize the RNA for 5 minutes. Sections were then blocked with Peroxidizer 1 and Background Sniper before adding CD68 (diluted 1:300, Dako) overnight. Polink-2 Plus HRP Mouse (GBI Labs) was added according to the manufacturer's directions. ImmPact DAB was added to visualize macrophages before counterstaining with CAT Hematoxylin and bluing in TBST. A thin layer of Clear mount diluted 1:5 in DI water (Thermo Scientific) was added to the sections, allowed to dry, dipped in xylenes before mounting in permount.

**RNAscope TSA ELF[2]**

Intracellular viral RNA and virions were visualized following the same protocol for the RNAscope 2.5 Red detection kit, but replacing Warp red staining with staining for 10 minutes using a 1:20 dilution of the ELF 97 Endogenous Phosphatase substrate from Invitrogen. Nuclei were stained with DAPI before mounting in Aqua polymount (Polysciences, Inc).

**RNAscope/ Immunofluorescence**

The RNAscope 2.5 Brown detection kit used the same method as above, except the addition of the Pretreat 1 (H2O2) step. Following the amp 6 step, Opal 570 (Perkin Elmer) was added following company instructions. IL6/CD68 and Napsin A (Biocare Medical) were added overnight. The following day, Donkey anti-rabbit Alexa Fluor 488 and Donkey anti-mouse Alexa Fluor 647 was added for 45 minutes before adding DAPI and mounting in Aqua polymount.

**IL-6 Immunohistochemistry**

Antigen retrieval was performed using 0.05% citraconic anhydride for 30 sec at 122 degrees using a declocker (Biocare Medical). Slides with 5 micron sections were cooled, and sections were circled with a hydrophobic pen (vector laboratories). Peroxidizer 1 was added for 5 minutes before washing in 3x in TBST and blocking in Background sniper (Biocare Medical) for 30 minutes. Rabbit anti-Human IL-6 antibody (Proteintech) was added diluted 1:1500 in DiVinci green (Biocare Medical) for 1 hour. A rabbit IgG monoclonal isotype control (Abcam) was used for the negative control on the patient tissue. Sections were washed 3x in TBST before adding the Polink-2 Plus HRP Rabbit (GBI Labs) according to the manufacturer's directions. ImmPact DAB (Vector Laboratories) was added before counterstaining with CAT Hematoxlyin (Biocare Medical) and bluing in TBST. Sides were dehydrated and mounted in permount.

**Data availability statement**

Limited samples were provided to the University of Minnesota by the Department of Biomedical and Clinical Sciences Luigi Sacco, University of Milan. All data are available in the main text or the supplementary materials.

**METHODS REFERENCES**

## ACKNOWLEDGMENTS

We thank Dr. Carlo Parravicini for the scientific discussion and the insights, Bianca Ghisi for organizing the communication of the Italian collaborators and Steve Wietgrefe for technical discussion.

## AUTHOR CONTRIBUTION

T.W.S, A.T.H. N.R.K and L.S. conceived the study. A.T.H and L.S. wrote the paper. A.T.H. designed the in situ hybridization and image analysis of the tissues. J.L.A performed all the experiments and acquired the images with the support of G.W. and B.G. M.G., M.C. and A.L. provided the laboratory samples, clinical information and provided expertise and feedback. N.R.K. provided expertise and feedback.

## COMPETING INTERESTS DECLARATION

The authors declare no competing interests.

## ADDITIONAL INFORMATION

**Material and Correspondence should be addressed to:**

*Luca Schifanella*: schif184@umn.edu
*Ashley Haase*: haase001@umn.edu
*Timothy Schacker*: schac008@umn.edu

**FIGURES**

**Figure 1.**

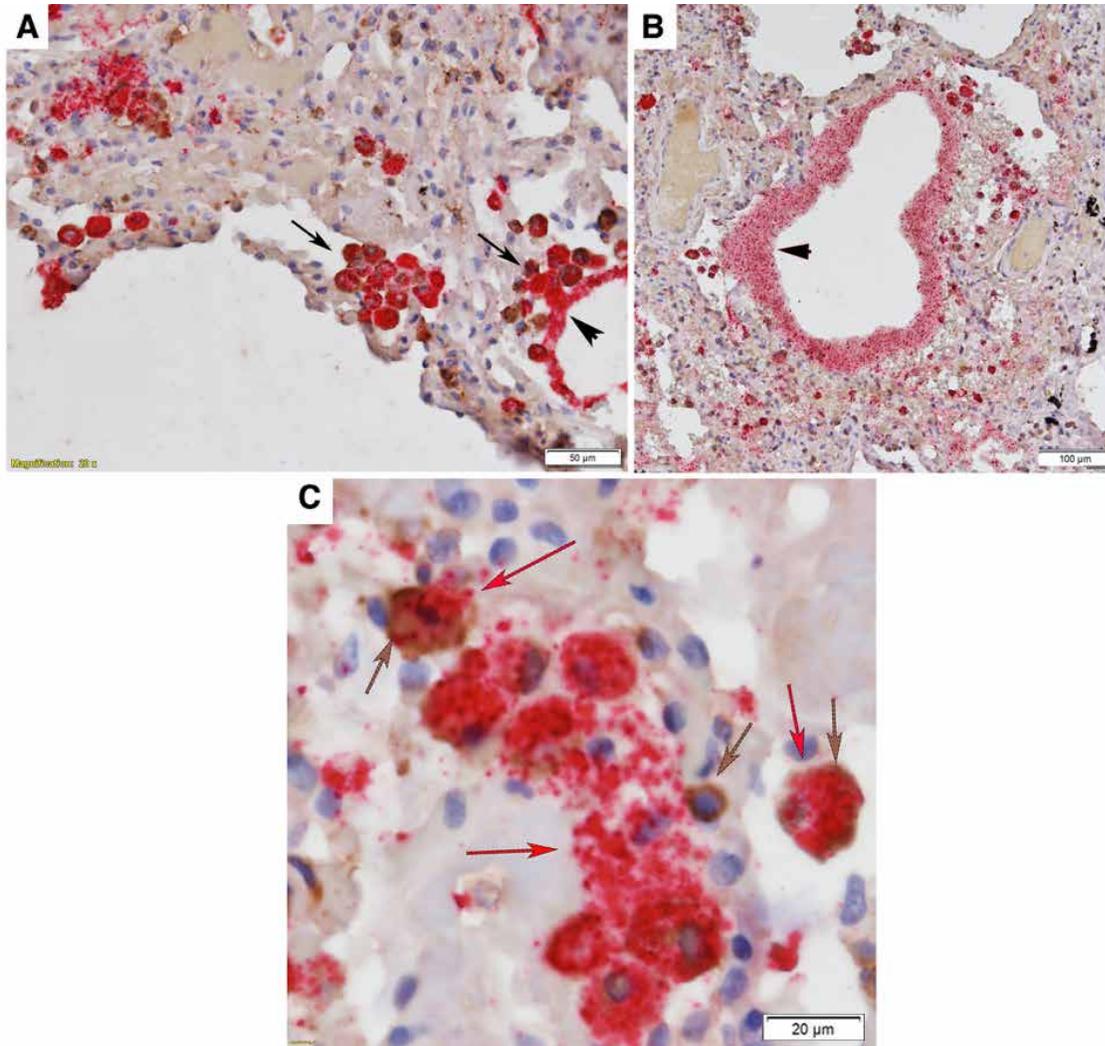

**Figure 1. SARS-CoV-2 replication and cytopathic effects in bronchial epithelium and type II pneumocytes. A, B.** Red viral RNA+ bronchial epithelium (short arrows) and clusters of viral RNA+ cells that by morphology appear to be hyperplastic type II pneumocytes (long arrows). **C.** Cluster of hyperplastic viral RNA+ epithelium. Red arrow points to viral RNA released from lysed cells. Single brown arrow points to a CD68+ macrophage. Red and brown arrows point to macrophages that appear to have phagocytosed degenerating infected epithelium.

**Figure 2.**

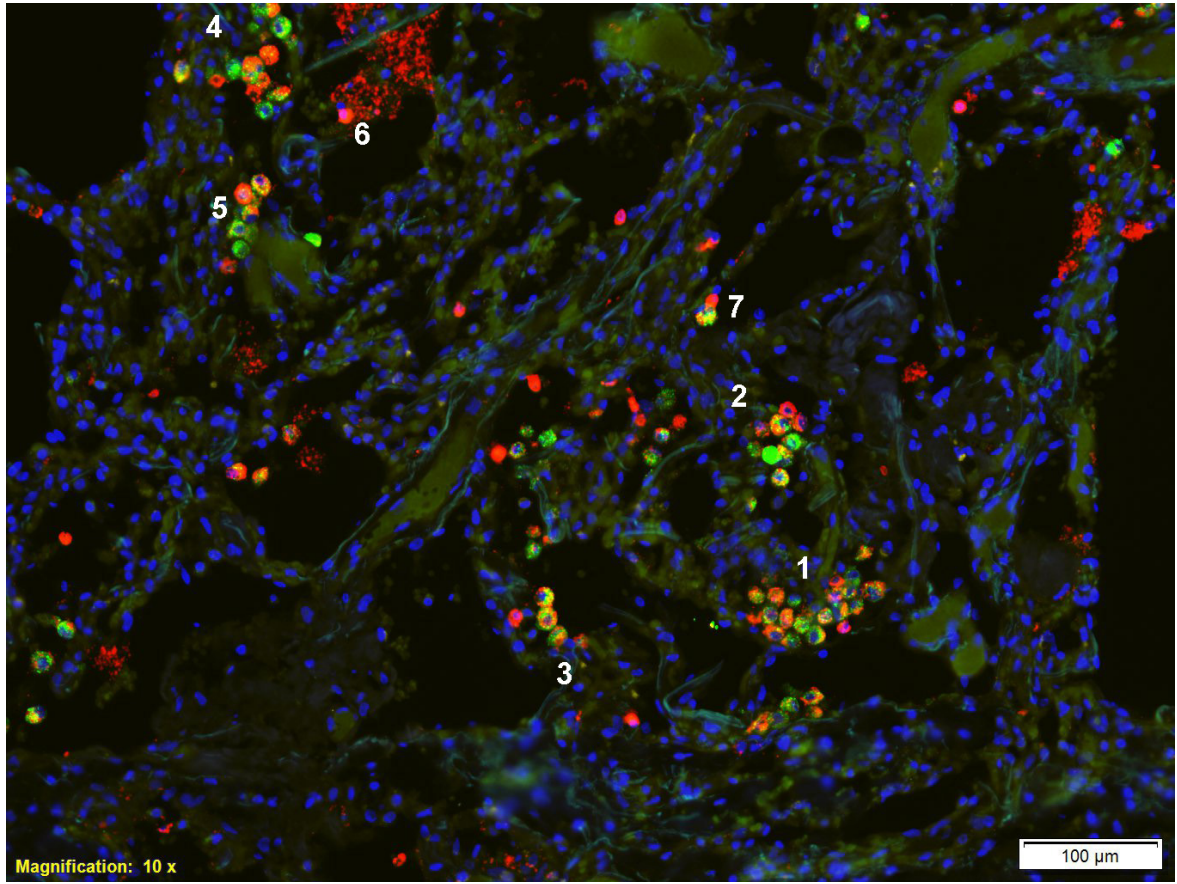

**Figure 2. Transmission chains in type II pneumocytes and interactions with other viral RNA+ cells.** Red cells are viral RNA+; green cells are Napsin+ type II pneumocytes; red-green cells are infected type II pneumocytes. The viral RNA+ type II cells in numbered regions 1-5 are spatially contiguous, consistent with spread of infection to susceptible type II cells in close spatial proximity. Region 6 shows a Napsin negative viral RNA+ cell overlying extracellular viral RNA+ derived from lysis of infected cells. Region 7 shows a viral RNA+ Napsin + and Napsin negative cell conjugate. These images are consistent with images in Figures 2 of acquisition of viral RNA in macrophages by phagocytosis.

**Figure 3.**

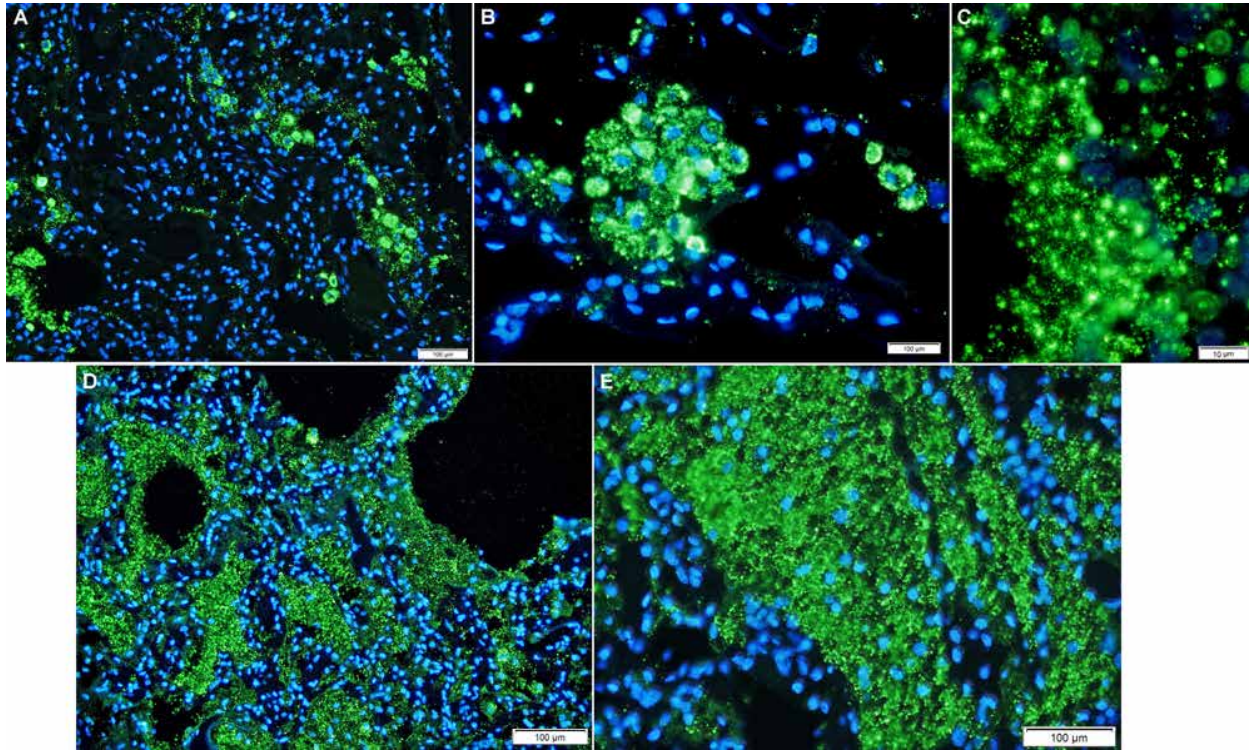

**Figure 3. Spread of infection in the lung by SARS-CoV-2 virus-producing cells and record of spread and cytopathic effects. A.** Green intracellular viral RNA, virions and aggregated virions are evident in individual cells, but also clusters and foci consistent with spread from cell-to-cell and to nearby susceptible cells. **B.** Cluster of virus-producing cells and three virus-producing cells in close spatial proximity. **C.** Individual virions, virus aggregates and inclusion bodies in and around dying cells. **D.** Overview of regions dominated by mats of lysed virus + cells in and adjacent to airways and around intact virus+ cells. **E.** Mat of virions in lysed type II pneumocytes

**Figure 4.**

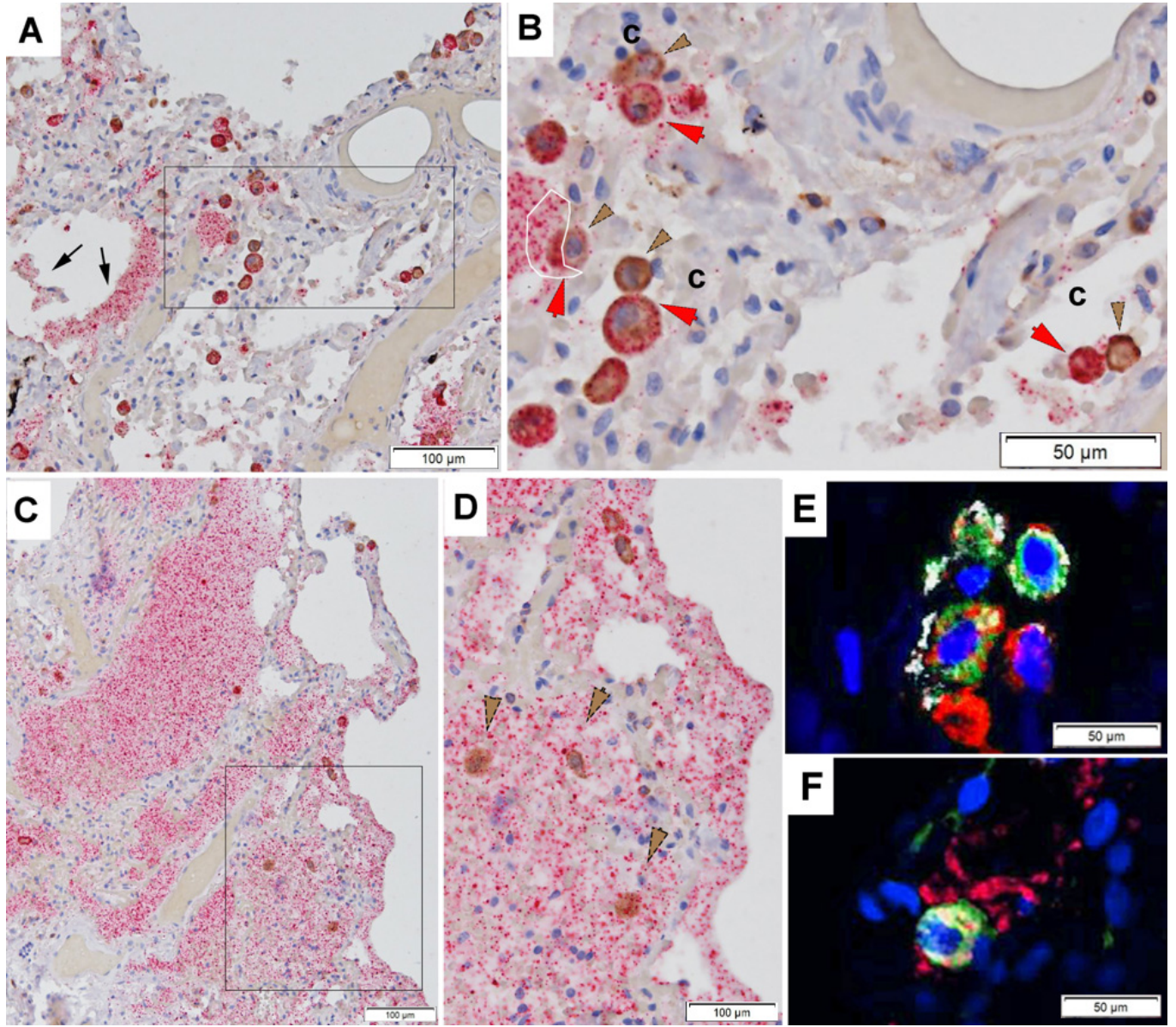

**Figure 4. Co-localization of macrophages with v+RNA+ cells and vRNA+ associated with lysed cells. A, B.** Arrows point to viral RNA+ in necrotic epithelium, a portion of which has been desquamated into alveolar space. Boxed region in A shown at higher magnification in B. Brown CD68+ viral RNA negative macrophages indicated by brown arrows; viral RNA+ cells indicated by red arrows. Conjugates of macrophages and viral RNA+ cells indicated by the letter c. The area enclosed by a white line and red arrow, delineate the viral RNA portion of a macrophage in contact with viral RNA from necrotic epithelium. **C, D.** Sheets of viral RNA+ epithelium. Region enclosed by a rectangle in C is shown at higher magnification in D. Brown arrows point to macrophages with particulate viral RNA+ signal. **E.** CD68+ (white) macrophages with viral RNA (red) associated with LAMP1+ (green) lysosomes. Extension of the macrophage, viral RNA and lysosomes co-localize with a viral RNA+ cell. Image taken at 40x magnification **F.** Macrophage contacting extracellular viral RNA. Viral RNA in the macrophage.

# EXTENDED DATA FIGURES

**Extended Data Figure 1.**

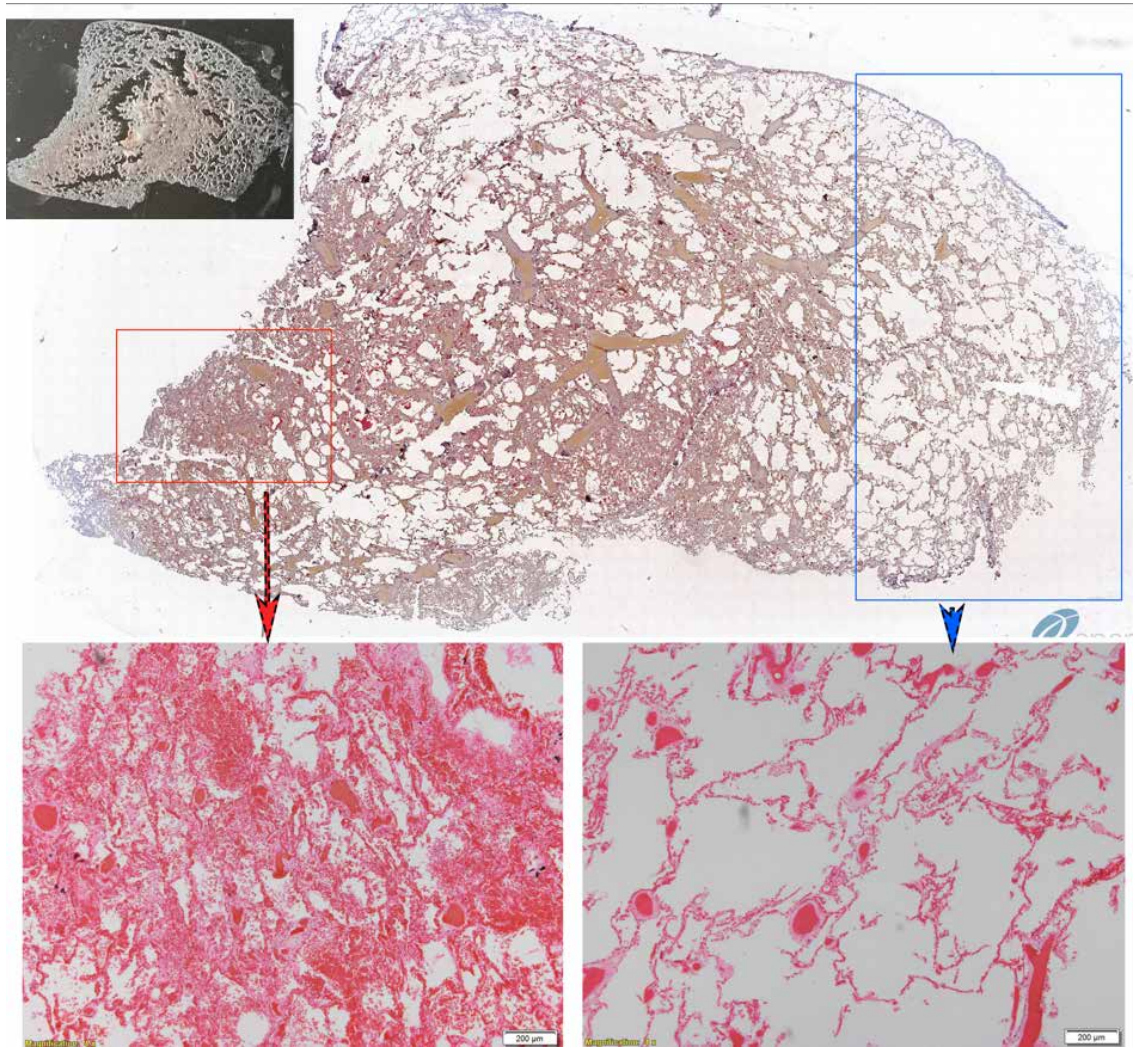

**Extended Data Figure 1. COVID-19 focal pneumonia.** Upper panel shows a thumbnail sketch and scan of lung tissue. Pathological findings in the brown-colored area enclosed by the red box. The red arrow points to the histopathology in H&E stained sections of diffuse alveolar damage with hyaline membranes, inflammatory cells, mainly lymphocytes, desquamated type II pneumocytes and red blood cells in the alveolar spaces. There were fewer pathological changes

in the region enclosed by the blue box. The blue arrow points to relative preservation of alveoli, although emphysematous changes were also noted.

**Extended Data Figure 2.**

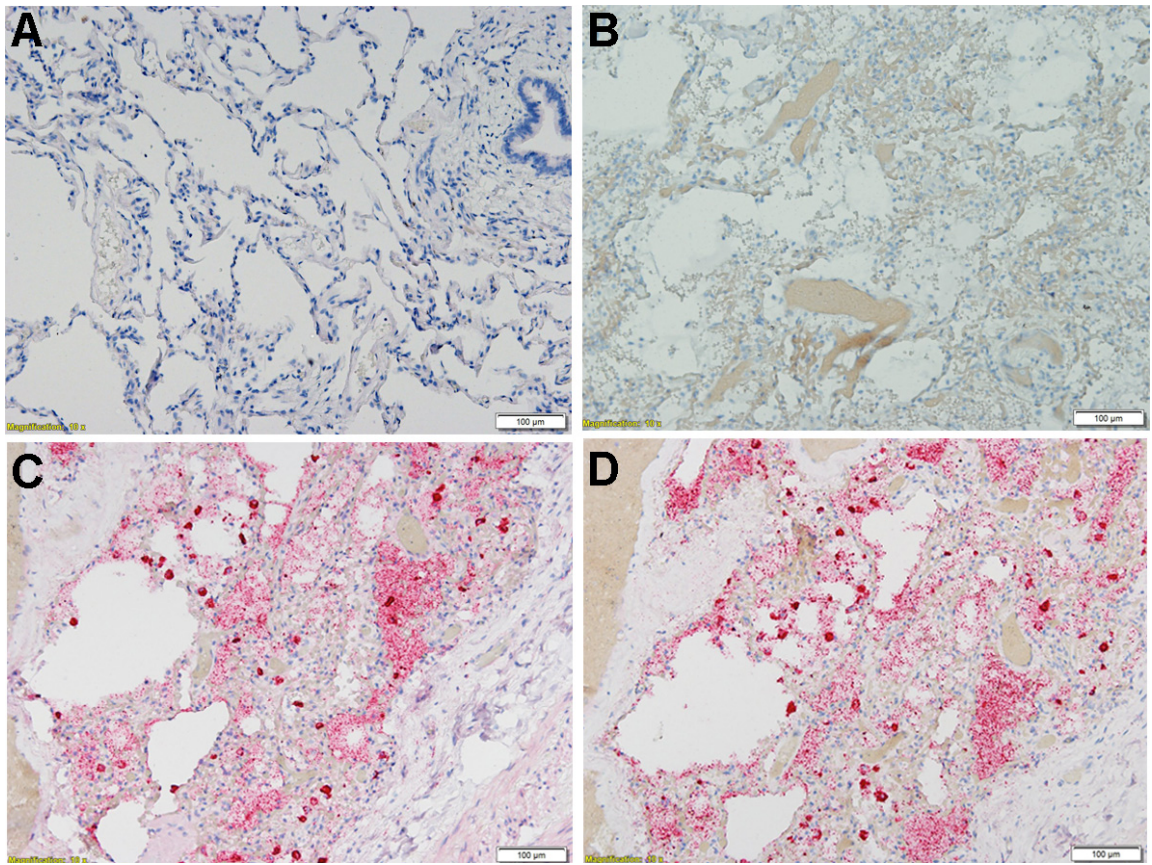

**Extended Data Figure 2. RNAscope in situ hybridization controls. A.** Normal lung hybridized with SN probes. **B.** Patient lung hybridized with a negative control probe (DapB, ACD). **C, D.** Comparable detection of genomic vRNA with only the SARS-CoV-2-specific S probe (**C**) and with only the N probe, which would also detect SARS-CoV and MERS-CoV genomic RNA (**D**).

**Extended Data Figure 3.**

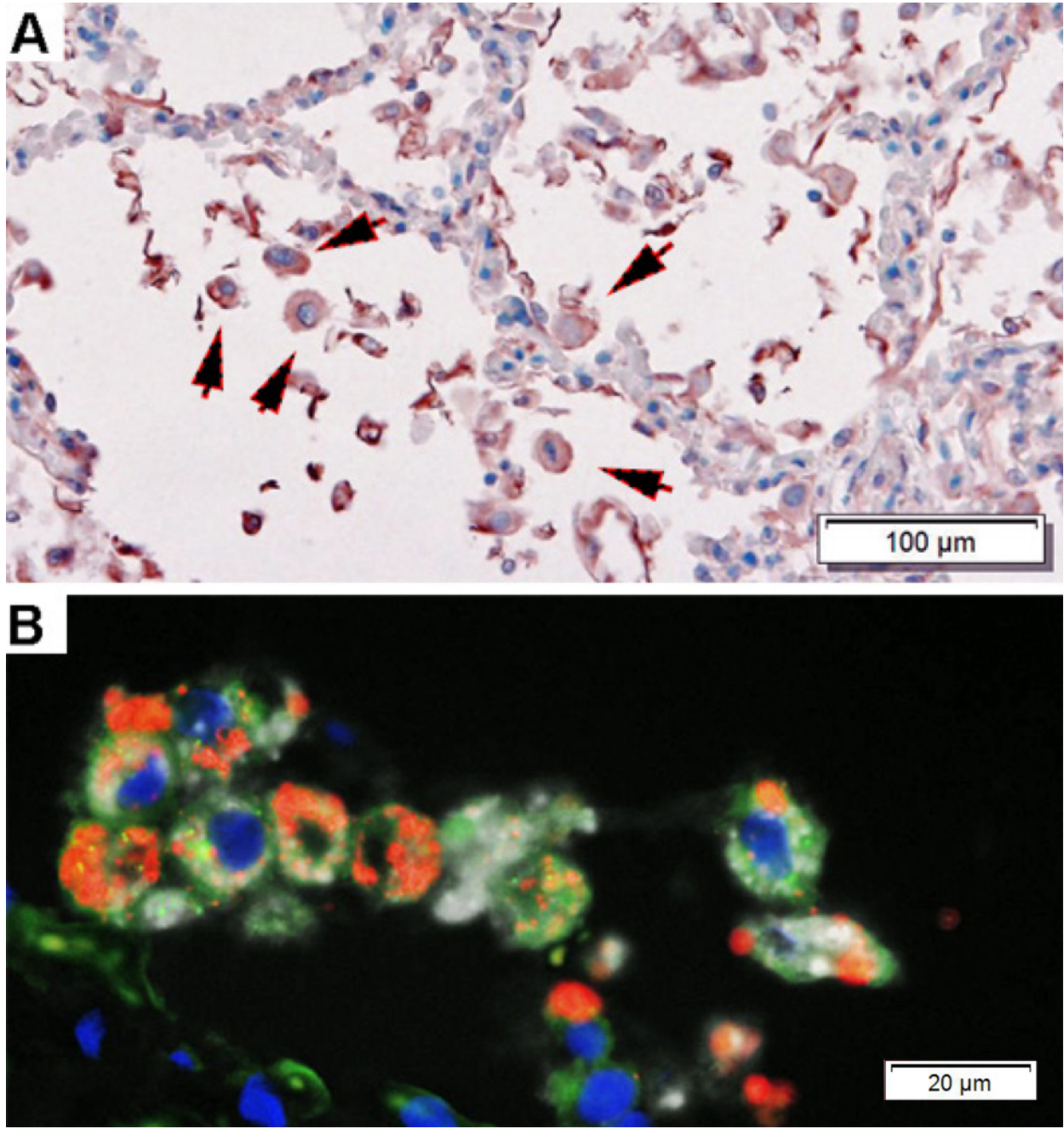

**Extended Data Figure 3. IL-6 production by type II pneumocytes. A.** Arrows point to brown IL-6 + hyperplastic infected type II cells. **B.** Green IL-6 + (green) Napsin + (white) type II pneumocytes with viral RNA (red).